\hsize=5.4in
\vsize=7in
\magnification=1200
\centerline{\bf Van der Waals force noise}
\bigskip
\centerline{Sh. Kogan}
\bigskip
\centerline{\it Los Alamos National Laboratory, Los Alamos, NM 87545}

\centerline{(Received 21 March 2005)}
\bigskip
The quantum theory of the fluctuations of the van der Waals (vdW) force between macroscopic bodies is developed. Unlike the mean vdW force that is determined by all quantum states that contribute to the optical absorption, the energies of those excitations of the interacting bodies that contribute effectively to the vdW force noise are thermal energies. Contrary to the mean vdW force the vdW force noise drops with decreasing temperature. Due to these differences the main mechanism of the mean vdW force and that of the vdW force noise may be different, e.g., the mean vdW force is determined by electronic excitations, the force noise by the random lattice or impurity dynamics. Since the vdW force is linear in the fields squared the dependence of the vdW force noise correlation function on time difference is determined not by one but by two frequencies: the spectral density depends not only on the frequency of its measurement $\omega_S$ but also on the integral over the second frequency. The dependence of its integrand on this frequency and temperature is quantum-like even at small $\omega_S\ll k_BT/\hbar$, and such may be the behaviour of the vdW noise. In the case of interacting metals the vdW force noise is enhanced by dc voltage applied across these metals. This additional noise is proportional to the voltage squared and to the spectral density of the random electric field at the frequency of noise measurement. The theory is in qualitative agreement with experiments.
\smallskip
\rightline{PACS numbers: 05.40.Ca, 07.79.Lh, 72.70.+m}
\bigskip
\centerline{\bf 1. INTRODUCTION}
\smallskip
After the fundamental works on Brownian motion by A. Einstein and M. Smoluchowski in the early 20-th century and after the experiments and theory on electrical noise in 1920s by J.B. Johnson and H. Nyquist it was realized that the ultimate limit of the sensitivity of any device and the accuracy of any measurement is determined by the fluctuations (noise) in the main physical phenomenon on which the device or measurement is based (not, for instance, by noise in the auxiliary circuitry). In the last two decades several types of scanning microscopy with atomic resolution have been developed: Scanning Tunneling Microscopy (STM), Scanning Atomic Force Microscopy (SAFM), and Magnetic Resonance Force Microscopy (MRFM). The Scanning Force Microscopies are based on the measurement of the force between the sample and the tip (probe) mounted on the end of a cantilever using a sensitive laser interferometry that measures the bending of the cantilever under the force. At present, owing to the progress of experimental methods, it is possible to measure very small forces on the order or even less than $10^{-18}$ N.$^{1-3}$  Recently the force noise in devices typical for SAFM have been measured.$^{4,5}$ The spectral density of the force noise was found to be $\sim 10^{-35}- 10^{-32}$ N$^2$/Hz.  The spectral density of the noise depends on the sample material (Ref. 5) and increases as the vacuum gap $L$ between the probe and the sample is decreased.$^{4,5}$ There are several questions that arise from these studies.

1. What is the minimum force measurable by nanomicroscopies? How does this ultimate limit depend on the materials, dimensions, and geometry of the tip and sample?

2. What is the temperature dependence of the van der Waals force noise?

3. Is the mechanism of the measured force noise inherent in the very mechanism of the measured force (and cannot be eliminated) or is it caused by extraneous mechanisms that can be eliminated by cleaning the surface of the sample, improving the cantilever quality and so on?

The main force between neutral bodies at distances between them in the range $\sim 1$ nm $-$ 1 $\mu$m is the van der Waals (below, for brevity, vdW) force. The vdW force between neutral atoms and molecules$^6$, atoms and lossless metals$^7$ or macroscopic bodies$^8$ is the result of correlation between the motion of charged particles in the interacting bodies. This correlation originates due to interaction via electromagnetic fields. In macroscopic bodies these fields are generated by random current (polarization) sources. E.M. Lifshitz calculated the mean vdW force between macroscopic bodies (per unit area) as the mean value of the Maxwell stress tensor, $T_{ij}$, in the vacuum gap near the surface of one of the interacting bodies (Ref. 8). The sources of Maxwell surface stress are all charges and currents inside the body enveloped by this surface:
$$T_{ij}({\bf r},t)= {1\over{4\pi}}\Bigl[\Bigl(-E_iE_j+{1\over2}\delta_{ij}E^2\Bigr)+\Bigl(-H_iH_j+ {1\over2}\delta_{ij}H^2\Bigr)\Bigr].\eqno(1.1)$$
Here and below ${\bf r}$ is the 2D radius vector parallel to the surface, ${\bf E}({\bf r},t)$ and ${\bf H}({\bf r},t)$ are the randomly varying electric and magnetic fields at the surface, respectively. In $T_{ij}$ the first index $i$ corresponds to the direction of the force, the second to the direction perpendicular to the element of the surface. In a realistic device  this Maxwell stress should be calculated for the surface of the tip (in Ref. 8 a mathematically simpler model was used). At a given distance $L$ between the probe's point and the sample surface (perpendicular to the $OX$ axis), the force acting on the probe in the $OX$ direction is an integral of the Maxwell stress $T_{xj}$ over the tip's surface, $A$:
$$F_x(t)=\int_A T_{xj}(t)dA_j,\eqno(1.2)$$
where ${\bf dA}$ is an element of the surface. For the system of two parallel semiinfinite slabs separated by a vacuum gap $L$, at temperature $T\to 0$ and small $L$, the mean vdW force per unit surface area reads$^8$:
$$\eqalign{\langle T_{xx}\rangle&={{\hbar}\over{4L^3}}\int_0^\infty{{d\xi}\over{2\pi}}\int_0^\infty{{dp}\over{2\pi}}{{p^2}\over{{{[\epsilon_1(i\xi)+1]}\over{[\epsilon_1(i\xi)-1]}}{{[\epsilon_2(i\xi)+1]}\over{[\epsilon_2(i\xi)-1]}}e^p-1}}\cr
&\approx {{\hbar}\over{8\pi^2L^3}}\int_0^\infty d\xi{{[\epsilon_1(i\xi)-1]}\over{[\epsilon_1(i\xi)+1]}}{{[\epsilon_2(i\xi)-1]}\over{[\epsilon_2(i\xi)+1]}}.\cr}\eqno(1.3)$$
Here $\xi$ is a real frequency variable. The dielectric permittivity at imaginary frequency, $\epsilon(i\xi)$, is a real quantity that monotonously falls off with $\xi$ from the value of the static dielectric permittivity $\epsilon(0)$ down to $\epsilon(i\infty)=1$, integration over dimensionless $p$ originates from the integration over the wave vector of the fields. Equation (1.3) is valid at small distances $L$ between the bodies when the retardation time of the electromagnetic waves traveling between the bodies can be neglected.$^8$ It happens when $L\ll c/\bar\omega$ where $c$ is the light speed and $\bar\omega$ is the maximum frequency of significant optical absorption. Practically, $\hbar \bar\omega\sim 10 - 100$ eV.

As follows from the theory$^{6,8}$, the mean vdW force between atoms or macroscopic bodies is determined by the entire spectrum of their optical absorption. The main part of this wide spectrum usually corresponds to frequencies $\omega\gg k_BT/\hbar$. At these frequencies, the quantum states that are the final states of the optical transitions are not populated. Therefore the variation of the temperature does not affect significantly the mean vdW force. We show below that the range of excitations' energies that determines the vdW force noise and its temperature dependence is drastically different from the energy range that determines the mean vdW force.

The force noise is measured by observing the random displacements of the cantilever. There are two mechanisms of random cantilever motion: 1) random motion of the mechanical system due to internal friction (damping) affected by the vdW interaction with the sample and 2) random variation of the electromagnetic fields that are the sources of the vdW force.
The results obtained in the first problem are presented and analyzed in the comprehensive review by Giessibl$^9$ (see also$^3$). This article is devoted to the second problem, i.e., to the theory of \lq\lq purely\rq\rq vdW noise.
\bigskip
\centerline{\bf 2. QUANTUM CORRELATION FUNCTIONS}
\smallskip
Let $F(t)$ be a fluctuating physical quantity, e.g. force, and $\langle F\rangle$ its mean value. Then $\delta F(t)=F(t)-\langle F\rangle$ is the time-dependent fluctuation of the quantity and $C(t_1,t_2)=\langle \delta F(t_1)\,\delta F(t_2)\rangle$ is the correlation function of the fluctuations (the angle brackets denote the statistical and quantum averaging). In general, $F(t)$ is a Heisenberg quantum-mechanical operator, $\hat F(t)$. Two such operators, $\delta \hat F(t_1)$ and $\delta \hat F(t_2)$, taken at different times $t_1\ne t_2$ (as in the correlation function), do not commute and $C(t_1,t_2)$ depends on their order. There are actually two correlation functions$^{10-12}$:
$$\eqalign{&C_-(t_1,t_2)\equiv C_-(t_1-t_2) =\langle \delta\hat F(t_1)\delta\hat F(t_2)\rangle=\langle \delta\hat F(t_1-t_2+t_0)\delta\hat F(t_0)\rangle, \cr
&C_+(t_1,t_2)\equiv C_+(t_1-t_2)=\langle \delta\hat F(t_2)\delta\hat F(t_1)\rangle=\langle \delta\hat F(t_0)\delta\hat F(t_1-t_2+t_0)\rangle =C_-[-(t_1-t_2)].\cr}\eqno(2.1)$$
According to the Wiener-Khintchine theorem, the spectral density is twice the Fourier transform of the correlation function. It follows that actually there are two Fourier transforms of the correlation functions and two spectral densities:
$$C_\pm(\omega)=\int_{-\infty}^{+\infty}dt e^{i\omega t}C_\pm(t),\qquad S_\pm(f)=2C_\pm (\omega)\quad t=t_1-t_2.\eqno(2.2)$$
Here $f=\omega/2\pi$ is the frequency assumed positive. Since $C_-(t)=C_+(-t)$, one can also present $S_-(f)$ and $C_-(\omega)$ as the functions $S_+(f)$ and $C_+(\omega)$ at the frequency with opposite sign:
$$S_-(f)=S_+(-f),\quad C_-(\omega)= C_+(-\omega).\eqno(2.2a)$$.

The correlation functions and spectral densities can be presented in terms of the eigenstates $|m\rangle$ of the system, their energies $E_m$, and the states' probabilities $w_m$ which for equilibrium systems are given by the Gibbs distribution:
$$\eqalign{&C_+(t)=\sum_m w_m\langle m|\delta\hat F(0)\delta\hat F(t)|m\rangle=\sum_{mn} w_m |(\delta F)_{mn}|^2 e^{i[(E_n-E_m)/\hbar]t},\cr
&C_-(t)=\sum_m w_m\langle m|\delta\hat F(t)\delta\hat F(0)|m\rangle=\sum_{mn} w_m |(\delta F)_{mn}|^2 e^{i[(E_m-E_n)/\hbar]t},\cr
&S_\pm(f)=2h\sum_{mn} w_m |(\delta\hat F)_{mn}|^2 \delta(E_m-E_n \mp hf).\cr}\eqno(2.3)$$
In an equilibrium system $w_m\propto \exp[-E_m/k_BT]$ and
$$S_-(f)=\exp[\hbar\omega/k_BT]S_+(f).\eqno(2.4)$$
The spectral density that corresponds to the commonly used symmetrized correlation function equals $S_s(f)={1\over2}[S_+(f)+S_-(f)]$. Using Eq. (2.4) one can easily express $S_{\pm}$ in terms of $S_s$ and vice versa.

The arguments of the delta functions in Eq. (2.3) show that $S_+(f)$ corresponds to the emission of energy from the sample (system) under study and $S_-(f)$ to the absorption of energy.$^{10-12}$ Each spectral density is proportional to the rate of emission or absorption under the condition that there is no backward reflection of the energy. The actually measured spectral density depends on the properties of the measuring device and the measurement conditions. If the sample under study is at absolute zero temperature it cannot radiate energy or momentum to the measuring device and the spectral density $S_+(f)=0$. However, if at the same time the device temperature $T_d>0$ the sample absorbs the noise energy emitted by the device, $S_-(f)\ne 0$. In principle, this spectrum also may be used as a characteristic noise spectrum of the sample.$^{10 - 12}$ If $T_d=0$ the measured \lq\lq absorption\rq\rq noise is also zero.

The spectral density $S_+(f)$ of the vdW force noise is zero at $T=0$ unlike the mean vdW force $\langle F\rangle$ that is non-zero at the same zero temperature (see Eq. (1.3)). This conclusion is in full agreement with the fact that at $T=0$ the system, being in the ground state, cannot transfer energy or momentum to the noise detector (see above). It means that the vdW force noise grows with temperature $T$. If the experiment actually measures $S_+(f)$ its value at $T=0$ is zero. The quantum theory of the force noise is instructive for understanding the fundamental difference between the properties of the mean vdW force and those of the vdW force noise.
\bigskip
\centerline{\bf 3. THE VAN DER WAALS FORCE NOISE}

\centerline{IN THE SYSTEM OF TWO SEMIINFINITE SLABS}
\smallskip
>From the mathematical point of view, the simplest model of a system of macroscopic bodies interacting via vdW force is the system of two semiinfinite slabs with parallel surfaces separated by a vacuum gap. It was introduced by E.M. Lifshitz$^8$ and later used by many authors. We shall use this model below to calculate the spectral densities $S_\pm(f)$ of fluctuating vdW surface stress and force. It is convenient in this model to introduce the 2D radius-vector ${\bf r}$ in the surface plane of the slabs and the $OX$ axis normal to this surface. The Langevin source in this problem is ${\bf K}(x,{\bf r},t)=4\pi {\bf P}(x,{\bf r},t)$, where ${\bf P}$ is the extraneous (source) polarization vector. In Ref. 8 the Fourier expansion of ${\bf P}$ in both slabs is introduced:
$$4\pi P_m(x,{\bf r},t)=\int d{\bf q}e^{i{\bf qr}}\int_0^{\infty}dk_x\cases{\cos(k_xx)g_m(k_x,{\bf q},t),&if $n=1$;\cr
\cos[k_x(x-L)]g_m(k_x,{\bf q},t),& if $n=2$.\cr}\eqno(3.1)$$
$$g_m(k_x,{\bf q},t)={2\over{\pi^2}}\int d{\bf r}\, e^{-i{\bf qr}}\int dx X_n(k_x;x)P_m(x,{\bf r},t),\eqno(3.2)$$
where
$$\int dx X_n(k_x;x)\ldots=\cases{\int_{-\infty}^0 dx\cos(k_x x)\ldots &if $n=1$;\cr
\int_L^\infty dx\cos[k_x(x-L)]\ldots &if $n=2$.\cr}\eqno(3.3)$$
In Eqs. (3.1) and (3.2) ${\bf q}$ is the wave vector in the plane of the surface, $i$ and $n$ number the vector components and the slabs, respectively, $m=(i,n)$, $g_m(k_x,{\bf q},t)$ is the cosine Fourier transform of the Langevin source in the $n$-th slab. As follows from Eq. (3.2) the Fourier amplitudes ${\bf g}_n(k_x,{\bf q},\omega)={\bf g}_n(-k_x,{\bf q},\omega)$, i.e., they are even functions of $k_x$.

We are using below the version of the Langevin approach used by Lifshitz$^8$. The electric field, ${\bf E}(x,{\bf r},t)$,  magnetic field ${\bf H}(x,{\bf r},t)$, the stresses $T_{ij}({\bf r},t)$ at the surfaces, and the Langevin sources of the fields are considered as quantum-mechanical operators satisfying the same equations as their classical counterparts (a comprehensive review of this approach is given by Henry and Kazarinov$^{13}$).

The correlation functions of the ${\bf g}_n$ sources can be expressed in terms of the polarization correlators. In general, the Fourier transform of the polarization correlator is proportional to the imaginary part of the dielectric permittivity $\epsilon_{ij}(x_1,{\bf r}_1;x_2,{\bf r}_2;\omega)$ in the linear integral relationship between the electric displacement $ D_i(x_1,{\bf r}_1,\omega)$ and the electric field $E_j(x_2,{\bf r}_2,\omega)$ (see$^{14}$). In the macroscopic approximation,
$$\epsilon_{ij}(x_1,{\bf r}_1;x_2,{\bf r}_2;\omega)\approx \delta(x_1-x_2)\delta({\bf r}_1-{\bf r}_2)\epsilon_{ij}(\omega).\eqno(3.4)$$
Due to $\delta(x_1-x_2)$ the double integral $\int dx_1X_n(k_{x1};x_1)\int dx_2X_n(k_{x2};x_2)$ in the correlator $\langle g_{m_1}(k_{x1},{\bf q}_1,\omega_1)g_{m_2}(k_{x2},{\bf q}_2,\omega_2)\rangle$ (Eq. (3.2))is reduced to one integral:
$$\int dx X_n(k_{x1};x)X_n(k_{x2};x) = {{\pi}\over 2}\delta(k_{x1}-k_{x2}).\eqno(3.5)$$

As follows from Eqs. (2.2),(3.2), (3.4), and (3.5) the correlators of the random quantities $g_{m}(k_{x},{\bf q},\omega)$ are:
$$\eqalign{&\langle g_{m_1}(k_{x1},{\bf q}_1,\omega_1)g_{m_2}(k_{x2},{\bf q}_2,\omega_2)\rangle
=2^4\delta(k_{x1}-k_{x2})\delta(\omega_1+\omega_2)\delta({\bf q}_1+{\bf q}_2)C_{-,m_1}(\omega_1)\cr
&\langle g^*_{m_1}(k_{x1},{\bf q}_1,\omega_1)g^*_{m_2}(k_{x2},{\bf q}_2,\omega_2)\rangle=2^4\delta(k_{x1}-k_{x2})\delta(\omega_1+\omega_2)\delta({\bf q}_1+{\bf q}_2)C_{-,m_1}(-\omega_1),\cr
&\langle g^*_{m_1}(k_{x1},{\bf q}_1,\omega_1)g_{m_2}(k_{x2},{\bf q}_2,\omega_2)\rangle=2^4\delta(k_{x1}-k_{x2})\delta(\omega_1-\omega_2)\delta({\bf q}_1-{\bf q}_2)C_{-,m_1}(-\omega_1),\cr
&\langle g_{m_1}(k_{x1},{\bf q}_1,\omega_1)g^*_{m_2}(k_{x2},{\bf q}_2,\omega_2)\rangle=2^4\delta(k_{x1}-k_{x2})\delta(\omega_1-\omega_2)\delta({\bf q}_1-{\bf q}_2)C_{-,m_1}(\omega_1).\cr}\eqno(3.6)$$
For brevity we omitted $\delta_{m_1m_2}$ in the rhs of these equations. Equations (3.6) are presented for positive frequency arguments of the functions $C_-$. If the frequency argument is negative this function is in fact a $C_+$ function (Eq. (2.2a)). The equilbrium correlation function of polarization is well known (see, for instance,$^{14}$ and Eq. (3.4)). For an isotropic solid,
$$\eqalign{&C_{\mp,m}(\omega)=\int d(x_1-x_2)d({\bf r}_1-{\bf r}_2)d(t_1-t_2)e^{\pm i\omega(t_1-t_2)}\langle P_m(x_1,{\bf r}_1,t_1)P_m(x_2,{\bf r}_2,t_2)\rangle\cr
&={{\hbar}\over{2\pi}}\epsilon''_n(\omega)\cases{N(\omega)+1&for $C_{-,m}(\omega)$;\cr
N(\omega)&for $C_{+,m}(\omega)$.\cr}.\cr}\eqno(3.7)$$
Here $N(\omega)$ is the Planck equation for the number of photons in one mode.

The vdW force is determined by the component $T_{xx}(x=0,{\bf r},t)$ of the Maxwell stress tensor:
$$T_{xx}({\bf r},t)={1\over{8\pi}}\bigl[E_y^2+E_z^2-E_x^2+H_y^2+H_z^2-H_x^2\bigr],\eqno(3.8)$$
where the fields are taken at $x=0$. The Fourier components of these fields are calculated in Ref. 8. At each given ${\bf q}$:
$$\eqalign{&E_x({\bf q},\omega)=- (q/p)(v_y-w_y),\quad E_y({\bf q},\omega)=v_y+w_y,\quad E_z({\bf q},\omega)= v_z+w_z,\cr
&H_x({\bf q},\omega)= {c\over{\omega}}q(v_z+w_z),\quad H_y({\bf q},\omega)=- {c\over{\omega}}p(v_z-w_z),\quad
H_z({\bf q},\omega)={{\omega}\over c}{1\over p}(v_y-w_y).\cr}\eqno(3.9)$$
Here $p=\sqrt{\omega^2/c^2-q^2}$, the vectors ${\bf v}$ and ${\bf w}$ depend on the wave vector ${\bf q}$ and the frequency $\omega$. They are integrals of the Langevin sources.$^8$ Introducing the notation
$$\tilde g_m({\bf q},\omega)=\int_{-\infty}^{+\infty}dk_x g_m(k_x,{\bf q},\omega)/(k_x^2-s_n^2),\eqno(3.10)$$
where $s_n^2=(\omega/c)^2\epsilon_n(\omega)-q^2$, we present them as follows:
$$\eqalign{&v_y+w_y={p\over{\Delta}}\Bigl\{s_1\Bigl[e^{-ipL}(\epsilon_2 p +s_2)-e^{ipL}(\epsilon_2 p-s_2)\Bigr][q\tilde g_{1x}-s_1\tilde g_{1y}]-2s_1s_2[q\tilde g_{2x}+s_2\tilde g_{2y}]\Bigr\},\cr
&v_y-w_y={p\over{\Delta}}\Bigl\{s_1\Bigl[e^{-ipL}(\epsilon_2 p +s_2)+e^{ipL}(\epsilon_2 p-s_2)\Bigr][q \tilde g_{1x}-s_1\tilde g_{1y}]+2s_2\epsilon_1p[q \tilde g_{2x}+s_2\tilde g_{2y}]\Bigr\},\cr
&v_z+w_z=-2p{{\omega^2}\over{c^2\Delta'}}[s_1e^{-ipL}\tilde g_{1z}+s_2\tilde g_{2z}],\quad
v_z-w_z=2{{\omega^2}\over{c^2\Delta'}}s_1s_2[-e^{-ipL}\tilde g_{1z}+\tilde g_{2z}],\cr
&\Delta=e^{ipL}(s_1-\epsilon_1 p)(s_2-\epsilon_2 p)-e^{-ipL}(s_1+\epsilon_1 p)(s_2+\epsilon_2 p),\cr
&\Delta'=e^{ipL}(s_1-p)(s_2- p)-e^{-ipL}(s_1+p)(s_2+p).\cr}\eqno(3.11)$$
The vectors ${\bf v}$ and ${\bf w}$ are linear in $\tilde g_m({\bf q},\omega)$. The coefficients do not depend on $k_x$.

As follows from Eqs. (3.8) and (3.9) the unaveraged stress at the surface $x=0$ equals:
$$\eqalign{&T_{xx}({\bf r},t)={1\over{8\pi}}\int {{d\omega d\omega'}\over{(2\pi)^2}}d{\bf q}d{\bf q}'e^{i[({\bf q}-{\bf q}'){\bf r}-(\omega-\omega')t]}\Bigl\{(v_y+w_y)({\bf q},\omega)(v_y+w_y)^*({\bf q}',\omega')\cr
&+[(\omega\omega'/c^2-qq')/pp'{^*}](v_y-w_y)({\bf q},\omega)(v_y-w_y)^*({\bf q}',\omega')\cr
&+[1-(c^2qq'/\omega\omega')](v_z+w_z)(v_z+w_z)^*({\bf q}',\omega')\cr
&+(c^2pp'{^*}/\omega\omega')(v_z-w_z)({\bf q},\omega)(v_z-w_z)^*({\bf q}',\omega')\Bigr\}.\cr}\eqno(3.12)$$

After substitution of Eqs. (3.11) in Eq. (3.12) we obtain the unaveraged stress $T_{xx}({\bf r},t)$ as a sum of terms each of which is proportional to a product $\tilde g_m({\bf q},\omega)\tilde g^*_{m'}({\bf q}',\omega')$. The equation for $T_{xx}$ takes the form:
$$T_{xx}({\bf r},t)={1\over{8\pi}}\int {{d{\bf q}d{\bf q}'d\omega d\omega'}\over{(2\pi)^2}}e^{i[({\bf q}-{\bf q}'){\bf r}-(\omega-\omega')t]}\sum_{m,m'}a_{mm'}({\bf q},\omega;{\bf q}',\omega')\tilde g_m({\bf q},\omega)\tilde g_{m'}^*({\bf q}',\omega').\eqno(3.13)$$
The coefficients $a_{mm'}$ are calculated in the Appendix using Eqs. (3.11) and (3.12). Since the stress must be real (Hermitian) these coefficients must satisfy the relationship:
$$a_{mm'}({\bf q},\omega;{\bf q}',\omega')=a^*_{m'm}({\bf q}',\omega';{\bf q},\omega).\eqno(3.14)$$
The unaveraged $T_{xx}$ (Eq. (3.12)) depends on $\omega$ and ${\bf q}$ in two ways: either directly on $\omega^2$ and $q^2$
or on arguments of the dielectric permittivities, $\epsilon_n({\bf q},\omega)$. It depends on the signs of ${\bf q}$ and $\omega$ only in the second way. Since $\epsilon_n(-{\bf q},-\omega)=\epsilon^*_n({\bf q},\omega)$ we obtain one more relationship:
$$a_{mm'}(-{\bf q},-\omega;-{\bf q}',-\omega')=a_{mm'}^*({\bf q},\omega;{\bf q}',\omega').\eqno(3.15)$$

Our goal is to calculate the correlation functions $C_{\pm,T_{xx}T_{xx}}({\bf r}_1,t_1;{\bf r}_2,t_2)$ and the corresponding spectral densities (Eq. (2.2)) of the surface stress fluctuations. Due to general relationship between the spectral densities $S_-(f)$ and $S_+(f)$ (Eq. (2.4)) the problem may be reduced to the calculation of $C_+$ and the corresponding $S_+(f)$:
$$C_{+,T_{xx}T_{xx}}({\bf r}_1,t_1;{\bf r}_2,t_2)=\langle T_{xx}({\bf r}_2,t_2)T_{xx}({\bf r}_1,t_1)\rangle-\langle T_{xx}(0,0)\rangle \langle T_{xx}(0,0)\rangle.\eqno(3.16)$$

Each correlation function is expressed in terms of the mean product of four random Langevin sources $\tilde g_m$. The Langevin sources in a macroscopic body are usually considered Gaussian. It follows that the mean value of this product can be decoupled and presented as a sum of products of pair correlation functions with the order of operators $\tilde g_m$ preserved within each pair. The correlation function equals:
$$\eqalign{&C_{+,T_{xx}T_{xx}}({\bf r}_1,t_1;{\bf r}_2,t_2)
=\Bigl({1\over{8\pi}}\Bigr)^2\int {{d{\bf q}_1d{\bf q}'_1d{\bf q}_2d{\bf q}'_2d\omega_1 d\omega'_1 d\omega_2 d\omega'_2}\over{(2\pi)^4}}\cr
&e^{i[({\bf q}_1-{\bf q}'_1){\bf r}_1+({\bf q}_2-{\bf q}'_2){\bf r}_2-(\omega_1-\omega'_1)t_1-(\omega_2-\omega'_2)t_2]}\cr
&\times\sum a_{m_1m'_1}({\bf q}_1,\omega_1;{\bf q}'_1,\omega'_1)a_{m_2m'_2}({\bf q}_2,\omega_2;{\bf q}'_2,\omega'_2)\cr
&\times\Bigl[\Big\langle \tilde g_{m_2}({\bf q}_2,\omega_2)\tilde g^*_{m'_2}({\bf q}'_2,\omega'_2)\tilde g_{m_1}(k_{x1},{\bf q}_1,\omega_1)\tilde g^*_{m'_1}({\bf q}'_1,\omega'_1)\Big\rangle\cr
&- \langle \tilde g_{m_2}({\bf q}_2\omega_2)\tilde g^*_{m'_2}({\bf q}'_2,\omega'_2)\rangle \,\langle \tilde g_{m_1}({\bf q}_1,\omega_1)\tilde g^*_{m'_1}({\bf q}'_1,\omega'_1)\rangle\Bigr].\cr}\eqno(3.17)$$
After decoupling, the expression in the rectangular brackets takes the form:
$$\eqalign{&\delta_{m_1m_2}\delta_{m'_1m'_2}\langle \tilde g_{m_1}({\bf q}_2,\omega_2)\tilde g_{m_1}({\bf q}_1,\omega_1)\rangle\, \langle \tilde g_{m'_1}^*({\bf q}'_2\omega'_2)\tilde g_{m'_1}^*({\bf q}'_1\omega'_1)\rangle \cr
&+\delta_{m'_1m_2}\delta_{m_1m'_2}\,\langle \tilde g_{m_1}^*({\bf q}'_2,\omega'_2)\tilde g_{m_1}({\bf q}_1,\omega_1)\rangle\,\langle \tilde g_{m'_1}({\bf q}_2,\omega_2)\tilde g_{m'_1}^*({\bf q}'_1,\omega'_1)\rangle.\cr}\eqno(3.18)$$

Using Eq. (3.6) the correlators of the $\tilde {\bf g}_n$ functions are expressed in terms of the correlators of the polarization sources of fluctuations:
$$\eqalign{&\langle \tilde g_m({\bf q}_2,\omega_2)\tilde g_m({\bf q}_1,\omega_1)\rangle=2^4\delta({\bf q}_1+{\bf q}_2)\delta(\omega_1+\omega_2)C_{+,m}(\omega_1)I_n(q_1,\omega_1),\cr
&\langle \tilde g_{m}^*({\bf q}'_2,\omega'_2)\tilde g_{m}^*({\bf q}'_1,\omega'_1)\rangle=2^4\delta({\bf q}'_1+{\bf q}'_2)\delta(\omega'_1+\omega'_2)C_{-,m}(\omega'_1)I_n(q'_1,\omega'_1),\cr
&\langle \tilde g_{m}^*({\bf q}'_2,\omega'_2)\tilde g_{m}({\bf q}_1,\omega_1)\rangle=2^4\delta({\bf q}_1-{\bf q}'_2)\delta(\omega_1-\omega'_2)C_{+,m}(\omega_1)I_n(q_1,\omega_1),\cr
&\langle \tilde g_{m}({\bf q}_2,\omega_2)\tilde g_{m}^*({\bf q}'_1,\omega'_1)\rangle=2^4\delta({\bf q}'_1-{\bf q}_2)\delta(\omega'_1-\omega_2)C_{-,m}(\omega'_1)I_n(q'_1,\omega'_1).\cr}\eqno(3.19)$$
Here
$$I_n(q,\omega)= \int_{-\infty}^{+\infty}{{dk_x}\over{[k_x^2-s_n^2(q,\omega)][k_x^2-s_n^2(q,-\omega)]}}={{\pi/2}\over{|s_n(q,\omega)|^2{\rm Im}\,s_n(q,\omega)}}.\eqno(3.20)$$

Thus the surface stress correlation function equals:
$$\eqalign{&C_{+,T_{xx}T_{xx}}({\bf r}_1,t_1;{\bf r}_2,t_2)={1\over{\pi^4}}\int d{\bf q}d{\bf q}'\int_{-\infty}^{+\infty}{{d\omega d\omega'}\over{(2\pi)^2}}e^{i[({\bf q}-{\bf q}')({\bf r}_1-{\bf r}_2)-(\omega-\omega')(t_1-t_2)]}\cr
&\times \sum_{mm'}a_{mm'}({\bf q},\omega;{\bf q}',\omega')\Bigl[[a_{mm'}(-{\bf q},-\omega;-{\bf q}',-\omega')+a_{m'm}({\bf q}',\omega';{\bf q},\omega)]\cr
&\times C_{+,m}({\bf q},\omega)C_{-,m'}(-{\bf q}',\omega')I_n(q,\omega)I_{n'}(q',\omega')\Bigr].\cr}\eqno(3.21)$$
According to Eqs. (3.14) and (3.15) the correlation function takes the form:
$$\eqalign{&C_{+,T_{xx}T_{xx}}({\bf r}_1,t_1;{\bf r}_2,t_2)={2\over{\pi^4}}\int d{\bf q}d{\bf q}'\int_{-\infty}^{+\infty}{{d\omega d\omega'}\over{(2\pi)^2}} e^{i[({\bf q}-{\bf q}')({\bf r}_1-{\bf r}_2)-(\omega-\omega')(t_1-t_2)]}\cr
&\times\sum_{mm'} |a_{mm'}({\bf q},\omega;{\bf q}',\omega')|^2C_{+,m}(\omega)C_{-,m'}(\omega')I_n(q,\omega)I_{n'}(q',\omega').\cr}\eqno(3.22)$$

The spectral density is twice the Fourier transform of the correlation function. Let $\omega_S$ be the frequency at which the spectral density is measured. After the substitution $\omega=\Omega+\omega_S/2$, $\omega'=\Omega-\omega_S/2$ the integration over frequencies is reduced to the integration over $\Omega$:
$$\eqalign{&S_{+,T_{xx}T_{xx}}({\bf r}_1-{\bf r}_2,\omega_S)\cr
&={2\over{\pi^4}}\int_{-\infty}^{+\infty}{{d\Omega}\over{2\pi}} \int d{\bf q}d{\bf q}'e^{i({\bf q}-{\bf q}')({\bf r}_1-{\bf r}_2)}\sum_{mm'} |a_{mm'}({\bf q},\Omega+\omega_S/2;{\bf q}',\Omega-\omega_S/2)|^2\cr
&\times C_{+,m}(\Omega+\omega_S/2)C_{-,m'}(\Omega-\omega_S/2)I_n(q,\Omega+\omega_S/2)I_{n'}(q',\Omega-\omega_S/2).\cr}\eqno(3.23)$$
The only temperature dependent quantities in Eq. (3.23) are the correlation functions $C_\pm$.

Using the relation $C_{\pm,m}(-\omega)=C_{\mp,m}(\omega)$ (Eq. (2.2a)) and Eqs. (3.14),(3.15),(3.20) it is easy to show that the integrand remains invariant under substitutions $\Omega\to -\Omega$, ${\bf q}\longleftrightarrow {\bf q}'$. Therefore we can integrate over positive frequencies $\Omega$ only:
$$\eqalign{&S_{+,T_{xx}T_{xx}}({\bf r}_1-{\bf r}_2,\omega_S)={4\over{\pi^4}}\int_0^\infty {{d\Omega}\over{2\pi}} \int d{\bf q}d{\bf q}'e^{i({\bf q}-{\bf q}')({\bf r}_1-{\bf r}_2)}\cr
&\times \sum_{mm'} |a_{mm'}({\bf q},\Omega+\omega_S/2;{\bf q}',\Omega-\omega_S/2)|^2I_n(q,\Omega+\omega_S/2)I_{n'}(q',\Omega-\omega_S/2)\cr
&\times C_{+,m}(\Omega+\omega_S/2)C_{-,m'}(\Omega-\omega_S/2).\cr}\eqno(3.24)$$

In experiments, not the stress but the force acting onto an area of the body is measured: $\hat F(t)=\int d{\bf r}T_{xx}({\bf r},t)$ (see Eq. (1.2)). $T_{xx}$ depends on ${\bf r}$ through the exponential function $e^{i({\bf q}-{\bf q}'){\bf r}}$ in the integrand. For a circular area of radius $r_c$ the integration over ${\bf r}$ replaces this function with $b=2\pi r_cJ_1(|{\bf q}-{\bf q}'|r_c)/|{\bf q}-{\bf q}'|$. The spectral density of this force, $S_F(\omega_S)$, is given by Eq. (3.24) in which the exponential function of ${\bf r}_1-{\bf r}_2$ is replaced by $b^2$:
$$\eqalign{&S_F(\omega_S)={8\over{\pi^3}}\int_0^\infty {{d\Omega}\over{2\pi}} \int d{\bf q}d{\bf q}'[r_cJ_1(|{\bf q}-{\bf q}'|r_c)/|{\bf q}-{\bf q}'|]^2\cr
&\times \sum_{mm'} |a_{mm'}({\bf q},\Omega+\omega_S/2;{\bf q}',\Omega-\omega_S/2)|^2 I_n(q,\Omega+\omega_S/2)I_{n'}(q',\Omega-\omega_S/2)\cr
&\times C_{+,m}(\Omega+\omega_S/2)C_{-,m'}(\Omega-\omega_S/2).\cr}\eqno(3.25)$$
The integration in Eqs. (3.24) and (3.25) is simplified owing to the dependence of the correlation functions only on $\Omega$ while other factors in the integrand, at $\Omega/c\ll L^{-1}$, depend, in fact, only on $q$ and $q'$.

The integrand in the equations for the spectral densities of $T_{xx}$ and vdW force is proportional to a product of {\it two} Langevin sources' spectral densities: one of $C_+$ type with frequency argument $\Omega+\omega_S/2$, the other of $C_-$ type with frequency $\Omega-\omega_S/2$. In that part of the integrand where $\Omega<\omega_S/2$ the frequency argument of $C_-$ is negative, i.e., according to Eq. (2.2a), both functions are of $C_+$ type and both tend to zero as $T\to 0$ (Eq. (2.3)). However in known experiments the frequency $\omega_S$ at which the spectral density is measured ($\sim 10^4 - 10^5$ Hz) is much smaller than the frequencies of strong dissipation $\epsilon''_n(\omega)$ because, in general, $\epsilon''(0)=0$. The main contribution to the force noise comes from $\Omega\gg \omega_S$. Then the temperature dependence of the vdW spectral density is determined by the product $C_+(\Omega)C_-(\Omega) \propto N(\Omega)[N(\Omega)+1]$. Due to $N(\Omega)$, the contribution of the dissipation $\epsilon''(\Omega)$ at $\Omega\gg k_BT/\hbar$ is exponentially small and does not contribute to the vdW force noise despite it fully contributes to the mean vdW force. Thus each term in the integrand of $S_{+,T_{xx}T_{xx}}$ falls off to zero as the temperature $T\to 0$ unlike the mean vdW force that tends to a non-zero value (Eq. (1.3)). An interesting peculiarity of the vdW force noise is also worth mentioning: even at small frequencies $\omega_S\ll k_BT/\hbar$, i.e. at conditions that are commonly considered as classical, the temperature dependence of the noise, due to the function $C_+(\Omega+\omega_S/2)$ in the integrand, behaves, at $\Omega> k_BT/\hbar$, as that of a quantum noise. However, if at $\omega_S<\Omega<\sim k_BT/\hbar$ the dissipation $\epsilon''_n(\Omega)$ is significant the temperature dependence of the force noise is $\sim T^2$.

The analysis presented above shows that only the excitations with energies $\hbar\omega$ smaller or on the order of thermal energy $k_BT$ effectively contribute to the vdW force noise. In this respect the vdW force noise differs from the mean vdW force which is determined by {\it all} excitations that contribute significantly to the electromagnetic dissipation in the interacting bodies, regardless of their energy in units $k_BT$ (Eq. (1.3)). Due to the fact that the contribution of the thermal excitation energies $\sim k_BT$ to the mean vdW force is usually insignificant and, on the contrary, only these energies contribute significantly to the vdW force noise the mechanisms of these two vdW phenomena may be quite different. The mechanism of the mean vdW force is mainly connected with electronic transitions, that of vdW force noise may result  mainly from lattice vibrations, e.g., transverse optical phonons, from the motion of impurities or from quasi-stationary currents in metals. The two measured quantities, the mean force and the force noise, supplement each other.

Other factors in Eqs. (3.24) and (3.25) (besides $C_{\pm,m}$) depend on $\Omega/c$. When $\Omega/c\ll L^{-1}$ the ratio
$\Omega/c$ can be dropped and these factors depend on $q$ and $q'$ only. Then (see Ref. 8),
$$p\approx iq; s_n\approx p; e^{\pm ipL}=e^{\mp qL}; I_n(q,\omega)I_{n'}(q',\omega')\propto (qq')^{-3}.\eqno(3.26)$$
The elements of the matrix $a_{mm'}({\bf q},\omega;{\bf q}',\omega')$ are calculated in the Appendix. In the approximation of small $(\Omega L/c)^2$, the elements $|a_{ni,n'i'}|^2$ with $i,i'=x,y$ are proportional to $(qq')^4$ and $a_{zn,zn'}\propto (\Omega/c)^2$ can be dropped. Then the part of the integrand in Eq. (3.25) that depends on $q$ and $q'$ can be presented as a function of $qL$ and $q'L$. One can see that if $r_c\sim L$ the spectral density of the vdW force is $\propto L^{-2}$.

The magnitude of the vdW force noise is determined, like that of the mean vdW force, by the complex dielectric permittivities of the interacting bodies. As is well known, in heteropolar dielectric crystals a strong IR absorption peak at the frequency of the transverse optical phonons, $\omega_t$, exists. In the range of this peak:
$$\epsilon''(\omega)\approx {{\pi}\over 2}\omega_t (\epsilon_0-\epsilon_\infty)\delta(\omega-\omega_t).\eqno(3.27)$$
The frequency $\omega_t$ is $\sim 10^{14}$ s$^{-1}$, i.e., $\hbar\omega_t\sim k_BT$ at room temperature $T$. This dissipation is a potential mechanism of significant vdW force noise. Substituting $[\epsilon''(\Omega)]^2$ in Eq. (3.25) we must replace the $\delta$-function peak with some more realistic, e.g. Lorentzian peak of finite width $\gamma$. In Eq. (3.25) and preceeding equations, the integral over the peak squared yields:
$$\int d\Omega[\epsilon''(\Omega)]^2\simeq{{\pi \omega_t^2(\epsilon_0-\epsilon_\infty)^2}\over{16\gamma}},\eqno(3.28)$$
Then the estimate of the force spectral density at small $\omega_S$ is:
$$S_{\pm,F}(\omega_S)={{[\hbar\omega_t\,(\epsilon_0-\epsilon_\infty)]^2)}\over{\gamma L^2}}{1\over{\exp[(\hbar\omega_t/k_BT)-1}}{1\over{1-\exp[(\hbar\omega_t/k_BT)}}.\eqno(3.29)$$
Substitution of characteristic values, $\epsilon_0-\epsilon_\infty\sim 1$, $L\sim 1$ nm, $\gamma\sim 0.1 \omega_t$, yields:
$$S_{\pm,F}(\omega_S)\sim {1\over{\exp[(\hbar\omega_t/k_BT)-1}}{1\over{1-\exp[-(\hbar\omega_t/k_BT)}}\cdot 10^{-35}{{{\rm N}^2}\over{{\rm Hz}}}.\eqno(3.30)$$
\bigskip
\centerline{\bf 4. EFFECT OF dc VOLTAGE BETWEEN MACROSCOPIC METALS}

\centerline{\bf ON THE VAN DER WAALS FORCE NOISE}
\smallskip

Consider a system of a probe and a sample that are both metallic or metallized. In this case a dc voltage $U$ can be                                                                                                                                                                                                                                                           applied across the vacuum gap between the probe and the sample. The goal is to find the effect of the dc voltage on the mean force and on the force noise. The mean stress (and force) is simply a sum of vdW stress $\langle T_{ij}\rangle_{\rm vdW}$ and the electrostatic stress due to the applied dc voltage, $(T_{ij})_{\rm U}$. The latter is proportional to $U^2$. These two forces are simply additive. However, the vdW force noise is strongly affected by the application of the dc voltage.$^5$

Let us calculate the change of the stress correlation function with voltage. The total electric field at any point ${\bf r}$ in vacuum is a sum of the dc field ${\bf F}({\bf r})$ and the randomly fluctuating field ${\bf E}({\bf r},t)$:
$${\bf E}_{\rm tot}({\bf r},t)={\bf F}({\bf r})+{\bf E}({\bf r},t).\eqno(4.1)$$
The total stress is a sum of three terms:
$$T_{ij}({\bf r},t)=T_{ij}^{(0)}({\bf r},t)+T_{ij}^{(1)}({\bf r},t)+T_{ij}^{(2)}({\bf r},t).\eqno(4.2)$$
Here the upper index is the total power of random field components in the term. As follows from Eq. (1.1),
$$T_{ij}^{(0)}={1\over{4\pi}}\bigl({1\over 2}\delta_{ij}F^2-F_iF_j\bigr),\quad T_{ij}^{(1)}={1\over{4\pi}}\bigl(\delta_{ij}{\bf FE}-F_iE_j-F_jE_i\bigr),\eqno(4.3)$$
and $T_{ij}^{(2)}$ is the stress due to the random electric and magnetic fields only. The correlation function of the total stress,
$$C_{T_{ij}T_{kl}}({\bf r}_1,t_1;{\bf r}_2,t_2)=\Big\langle T_{ij}({\bf r}_1,t_1)T_{kl}({\bf r}_2,t_2)\Big\rangle -
\langle T_{ij}({\bf r}_1,t_1)\rangle\,\langle T_{kl}({\bf r}_2,t_2)\rangle.\eqno(4.4)$$
We substitute Eq. (4.2) in this equation and take into account that in equilibrium the mean value of products of odd number of random fields is zero. The variation of the correlation function due to the static electric field (dc voltage $U$) equals:
$$\eqalign{&C^{(U)}_{T_{ij}T_{kl}}({\bf r}_1,t_1;{\bf r}_2,t_2)=\Big\langle T_{ij}^{(1)}({\bf r}_1,t_1)T_{kl}^{(1)}({\bf r}_2,t_2)\Big \rangle\cr
&={1\over{(4\pi)^2}}\Big\langle\Bigl(\delta_{ij}{\bf F}({\bf r}_1)\cdot{\bf E}({\bf r}_1,t_1)-F_i({\bf r}_1)E_j({\bf r}_1,t_1)-F_j({\bf r}_1)E_i({\bf r}_1,t_1)\Bigr)\cr
&\times\Bigl(\delta_{kl}{\bf F}({\bf r}_2)\cdot{\bf E}({\bf r}_2,t_2)-F_k({\bf r}_2)E_l({\bf r}_2,t_2)-F_l({\bf r}_2)E_k({\bf r}_2,t_2)\Bigr)\Big\rangle.\cr}\eqno(4.5)$$
Obviously, at any geometry of the probe and sample, the spectral density of the additional noise due to dc voltage, $S^{(U)}(f)$, is proportional to  $U^2$, does not depend on the sign of the voltage, and is linear in the spectral densities of the random electric field ${\bf E}$ components, $S_{E_iE_j}(f;{\bf r}_1-{\bf r}_2)$, at the frequency of the spectral density of noise measurement (no additional integration over frequency). It means that the electric field spectral densities  determine the steepness of the additional noise growth with $U^2$. At \lq\lq classical\rq\rq frequencies $f\ll k_BT/h$ the spectral densities $S_{E_iE_j}(f)$ are roughly linear in temperature $T$. Under these conditions the noise $S^{(U)}(f)$ grows linearly with temperature. It must be emphasized that the noise $\propto U^2$ is created by a noiseless dc voltage and therefore can be viewed as an amplification of noise generated by the only its sources, i.e., by the sources of the random electromagnetic fields.

In the model of two semiinfinite parallel metals separated by a vacuum gap of thickness $L$ the dc field in vacuum $F_x=(U/L)$, and the vdW force is determined only by $T_{xx}$. It follows from Eq. (4.5) that the spectral density of the stress equals:
$$S^{(U)}_{T_{xx}T_{xx}}(f;{\bf r}_1-{\bf r}_2)={{U^2}\over{L^2}}S_{E_xE_x}(f;{\bf r}_1-{\bf r}_2).\eqno(4.6)$$
The problem has been reduced to the calculation of the spectral density of the electric field $E_x$ in the vacuum gap near the surface of any of the two slabs.
\bigskip
\centerline{\bf 5. MAIN CONCLUSIONS AND COMPARISON WITH EXPERIMENTS}
\smallskip
Since the geometry of the theoretical model (two semiinfinite slabs) strongly differs from that of the real SAFM device (acute metallic tip above the sample) the quantitative comparison of the theory and experiment is not justified. Therefore we present below a summary of qualitative results and compare them with the known experimental data.

First, we found that the properties of the mean van der Waals force and the wdW force noise strongly differ. The former are determined by all states of the interacting bodies that contribute to the optical transitions in these bodies. At common temperatures of measurement, the major part of these quantum states are not populated. The mean vdW force is therefore non-zero at zero absolute temperature and the variation of the temperature only weakly affects this zero-temperature value. On the contrary, the measurement of the force noise requires transfer of energy and/or momentum from the noisy system to the noise detector. Only the quantum states with energies (reckoned from the ground state) less or on the order of $k_BT$ contribute to the force noise. That is why the force noise, unlike the mean force, falls off at $T\to 0$. At low temperatures it falls off exponentially. This conclusion is qualitatively supported by experimental findings$^5$: in the system Au film$-$Au-coated cantilever the effective damping coefficient $\Gamma=S_F/4k_BT$  was found to be $\sim$6 times lower at 77 K than at 300 K. It means that the spectral density of the force noise was $\sim$24 times smaller at temperature only 4 times lower.

Van der Waals forces between two bodies are due to the correlation of charges and currents in both bodies via electromagnetic fields. Since the system's excitations with energies $<\sim k_BT$ comprise a very small contribution to the mean vdW force and, on the contrary, they comprise the main contribution to the vdW force noise, the mechanisms of the electromagnetic fields' sources, i.e. the origin of the vdW forces may be quite different in both cases.

The vdW force between macroscopic bodies is expressed in terms of the Maxwell surface stress, $T_{ij}({\bf r},t)$, that is bilinear in the components of the randomly varying electric and magnetic fields. Therefore the spectral density of the vdW force noise depends on the (given) frequency $\omega_S$ at which it is measured but is also an integral over the second frequency, $\Omega$ (Sec. 3). The integrand is proportional to the {\it product of two} spectral densities of the random electromagnetic fields' Langevin sources. In the range of small $\omega_S$, much smaller than the frequencies of significant dissipation ($\epsilon''_{1,2}$), the spectral density of the vdW force noise is practically independent of $\omega_S$. Then the temperature dependence of the force noise is determined by the integral over $\Omega$ the integrand of which depends on $\exp[\hbar\Omega/k_BT]$. If the frequencies of strong dissipation are $> k_BT/\hbar$ the noise is quantum even at $\omega_S\ll k_BT/\hbar$.

In experiments the dependence of the vdW noise on $L$ is $\sim L^{-d}$ with $d=3$ (Ref. 4) and $d=1.3$ (Ref. 5), respectively (the difference is due to different geometry of the used systems \lq\lq cantilever$-$sample\rq\rq$^5$).  According to Sec. 3, in the two-slab model the force noise falls off with $L$ also as $\sim L^{-d}$ but with $d\simeq 2$. This dependence is close to the experimental ones.

The striking phenomenon of force noise enhancement between two metals by dc voltage$^5$ follows immediately from the general equation for the vdW force (Sec. 4). According to the experiment$^5$ and to the theory the additional force noise is proportional to the dc voltage squared. This agreement between the theory and experiment proves once more that both the force and force noise between the two metals are of vdW origin.

As is shown in Sec. 3, the strong IR absorption peak at the frequency of the transverse optical phonons, $\omega_t$, may be a significant source of the vdW noise. It was found experimentally$^5$ that the vdW force noise in quartz is much higher than in the case of metal samples. This difference can be attributed to the dissipation peak around $\omega_t$. The rough estimate (3.30) of the vdW force noise spectral density in the two slabs model does not strongly differ from the experimental data$^5$. Some part of this difference may be attributed to the difference between the simple geometry of the   two infinite slabs model and the geometry of the real experimental device.
\bigskip
\centerline{\bf ACKNOWLEDGEMENTS}
\smallskip
I am very grateful to Dr. G.P. Berman of the LANL Theoretical Division for many fruitful discussions. I also highly appreciate the interesting discussions with Dr. P. Milonni, his criticism, comments and remarks.  This work was partially financially supported by DARPA.
\vfill\eject
\centerline{APPENDIX: THE MATRIX $a_{mm'}$}

Let $A_{j,ni}$ be the coefficient of $\tilde g_{ni}$ in the $j$-th term of the four terms in Eq. (3.11): 1) $v_y+w_y$,
2) $v_y-w_y$, 3) $v_z+w_z$, 4) $v_z-w_z$. Comparing $a_{mm'}\tilde g_m \tilde g^*_{m'}$ and Eq. (3.11) we obtain:
$$\eqalign{&A_{1,1x}={{pqs_1}\over{\Delta}}\Bigl[(s_2-\epsilon_2 p)e^{ipL}+(s_2+\epsilon_2 p)e^{-ipL}\Bigr],\cr
&A_{2,1x}=-{{pqs_1}\over{\Delta}}\Bigl[(s_2-\epsilon_2 p)e^{ipL}-(s_2+\epsilon_2 p)e^{-ipL}\Bigr],\cr
&A_{1,1y}=-{{ps_1^2}\over{\Delta}}\Bigl[(s_2-\epsilon_2 p)e^{ipL}+(s_2+\epsilon_2 p)e^{-ipL}\Bigr],\cr
&A_{2,1y}={{ps_1^2}\over{\Delta}}\Bigl[(s_2-\epsilon_2 p)e^{ipL}-(s_2+\epsilon_2 p)e^{-ipL}\Bigr],\cr
&A_{1,2x}=-{{2pqs_1s_2}\over{\Delta}},\quad A_{2,2x}={{2p^2q\epsilon_1s_2}\over{\Delta}},\cr
& A_{1,2y}=-{{2ps_1s_2^2}\over{\Delta}},\quad A_{2,2y}={{2\epsilon_1p^2s_2^2}\over{\Delta}},\cr
&A_{3,1z}=-{{2ps_1\omega^2}\over{c^2\Delta'}}e^{-ipL},\quad A_{4,1z}=-{{2s_1s_2\omega^2}\over{c^2\Delta'}}e^{-ipL},\cr
&A_{3,2z}=-{{2ps_2\omega^2}\over{c^2\Delta'}},\quad A_{4,2z}={{2s_1s_2\omega^2}\over{c^2\Delta'}}.\cr}\eqno(A1)$$

The matrix elements $a_{mm'}$ are presented in terms of $A_{j,m}$:
$$\eqalign{&a_{1x,1x}=A_{1,1x}(q,\omega)A^*_{1,1x}(q',\omega')+{{\omega\omega'/c^2 -qq'}\over{p(q,\omega)p^*(q',\omega')}}
A_{2,1x}(q,\omega)A^*_{2,1x}(q',\omega'),\cr
&a_{1y,1y}=A_{1,1y}(q,\omega)A^*_{1,1y}(q',\omega')+{{\omega\omega'/c^2 -qq'}\over{p(q,\omega)p^*(q',\omega')}}
A_{2,1y}(q,\omega)A^*_{2,1y}(q',\omega'),\cr
&a_{1x,1y}=A_{1,1x}(q,\omega)A^*_{1,1y}(q',\omega')+{{\omega\omega'/c^2 -qq'}\over{p(q,\omega)p^*(q',\omega')}}
A_{2,1x}(q,\omega)A^*_{2,1y}(q',\omega'),\cr
&a_{1y,1x}=A_{1,1y}(q,\omega)A^*_{1,1x}(q',\omega')+{{\omega\omega'/c^2 -qq'}\over{p(q,\omega)p^*(q',\omega')}}
A_{2,1y}(q,\omega)A^*_{2,1x}(q',\omega'),\cr
&a_{1x,2x}=A_{1,1x}(q,\omega)A^*_{1,2x}(q',\omega')+{{\omega\omega'/c^2 -qq'}\over{p(q,\omega)p^*(q',\omega')}}
A_{2,1x}(q,\omega)A^*_{2,2x}(q',\omega'),\cr
&a_{1y,2y}=A_{1,1y}(q,\omega)A^*_{1,2y}(q',\omega')+{{\omega\omega'/c^2 -qq'}\over{p(q,\omega)p^*(q',\omega')}}
A_{2,1y}(q,\omega)A^*_{2,2y}(q',\omega'),\cr
&a_{1x,2y}=A_{1,1x}(q,\omega)A^*_{1,2y}(q',\omega')+{{\omega\omega'/c^2 -qq'}\over{p(q,\omega)p^*(q',\omega')}}
A_{2,1x}(q,\omega)A^*_{2,2y}(q',\omega'),\cr
&a_{1y,2x}=A_{1,1y}(q,\omega)A^*_{1,2x}(q',\omega')+{{\omega\omega'/c^2 -qq'}\over{p(q,\omega)p^*(q',\omega')}}
A_{2,1y}(q,\omega)A^*_{2,2x}(q',\omega').\cr}\eqno(A2)$$

$$\eqalign{&a_{1z,1z}=\bigl[1-(c^2/\omega\omega')qq'\bigr]A_{3,1z}(q,\omega)A^*_{3,1z}(q',\omega')\cr
&+(c^2/\omega\omega')p(q,\omega)p^*(q',\omega')A_{4,1z}(q,\omega)A^*_{4,1z}(q',\omega'),\cr
&a_{2z,2z}=\bigl[1-(c^2/\omega\omega')qq'\bigr]A_{3,2z}(q,\omega)A^*_{3,2z}(q',\omega')\cr
&+(c^2/\omega\omega')p(q,\omega)p^*(q',\omega')A_{4,2z}(q,\omega)A^*_{4,2z}(q',\omega'),\cr
&a_{1z,2z}=\bigl[1-(c^2/\omega\omega')qq'\bigr]A_{3,1z}(q,\omega)A^*_{3,2z}(q',\omega')\cr
&+(c^2/\omega\omega')p(q,\omega)p^*(q',\omega')A_{4,1z}(q,\omega)A^*_{4,2z}(q',\omega').\cr}\eqno(A2a)$$
\bigskip
\noindent e-mail: shkogan@comcast.net
\smallskip
\centerline{\bf REFERENCES}
\bigskip
$^1$ D. Rugar, B.C. Stipe, H.J. Mamin, C.S. Yannoni, T.D. Stowe, K.Y. Yasumura, and T.W. Kenny, Appl. Phys. A {\bf 72}, [Suppl.] (2001)

$^2$ H.J. Mamin and D. Rugar, Appl. Phys. Lett. {\bf 79}, 3358 (2001)

$^3$ D. Rugar, R. Budakian, B.W. Chui, and H.J. Mamin, Proc. of SPIE {\bf 5112} (2003)

$^4$ I. Dorofeyev, H. Fuchs, G. Wenning, and B. Gotsmann, Phys. Rev. Lett. {\bf 83}, 2402 (1999)

$^5$ B.C. Stipe, H.J. Mamin, T.D. Stowe, T.W. Kenny, and D. Rugar, Phys. Rev. Lett. {\bf 87}, 096801 (2001)

$^6$ F. London, Zs. f\"ur Physik {\bf 60}, 491 (1930)

$^7$ H.B.C. Casimir and D. Polder, Phys. Rev. {\bf 73}, 360 (1948)

$^8$ E.M. Lifshitz, Zh. Eksper. Teor. Phys.  {\bf 29}, 94 (1955)
[Soviet Phys. - JETP {\bf 2},73 (1956)]

$^9$ F.J. Giessibl, Rev. Mod. Phys. {\bf 75}, 949 (2003)

$^{10}$ G.B. Lesovik and R. Loosen, Pis'ma Zhur. Eksper. Teor. Fiz. {\bf 65}, 280 (1997)[JETP Lett. {\bf 65}, 295 (1997)]

$^{11}$ G.B. Lesovik, Uspekhi Fiz. Nauk {\bf 168}, 155 (1998) [Physics - Uspekhi {\bf 41}, 145 (1998)]

$^{12}$ U. Gavish, Y. Levinson, and Y. Imry, Phys. Rev. B {\bf 62}, 10637 (2000)

$^{13}$ C.H. Henry and R.F. Kazarinov, Rev. Modern Phys. {\bf 68}, 801 (1996)

$^{14}$ Sh. Kogan, {\it Electronic Noise and Fluctuations in Solids} (Cambridge University Press, 1996)

\end